\documentclass[aps,pra,twocolumn,a4paper,english]{revtex4-2}
\usepackage{amsmath,epsfig,amssymb,subfigure,bm,dsfont}
\usepackage{lipsum}
\usepackage{amssymb}
\usepackage{mathrsfs,amsbsy,color,graphicx,amsthm,amsfonts}
\usepackage{units}
\usepackage{bbm}

\begin{document}

\title{Temperature-related single-photon transport in waveguide QED}
\author{Wei-Bin Yan}
\affiliation{College of Physics and Engineering, Qufu Normal University, Qufu, 273165,
China}
\author{Zhong-Xiao Man}
\email{manzhongxiao@163.com}
\affiliation{College of Physics and Engineering, Qufu Normal University, Qufu, 273165,
China}
\author{Ying-Jie Zhang}
\affiliation{College of Physics and Engineering, Qufu Normal University, Qufu, 273165,
China}
\author{Yun-Jie Xia}
\affiliation{College of Physics and Engineering, Qufu Normal University, Qufu, 273165,
China}

\begin{abstract}
We propose a scheme to realize the single-photon transport affected by the
temperature. The scheme is composed by a waveguide-atom interacting
structure linked to a thermal bath. The single-photon reflection coefficient
can be tuned by adjusting the temperature of the
thermal bath. This provides a thermal control of the single-photon
transport. Moreover, the temperature of the thermal bath can be estimated by
measuring the single-photon transport. It is feasible that the estimation on
the temperature is sensitive to slight changes of low temperature. This
implies an avenue for implementing the optical thermometer with the ability to
accurately measure the sample temperature in the low-temperature region.
\end{abstract}

\maketitle

\section{Introduction}

Single photons are considered as the brilliant long-distance carrier of
quantum information. Control of the single-photon transport plays an
important role in quantum information processing. Waveguide QED \cite{Sheremet2023,Royd2017}, which has
attracted widespread attention, provides an excellent platform for the
controllable single-photon transport. Recently, the controllable single-photon
transport in waveguide QED has been extensively studied, such as the
single-photon switch \cite{2005Coherent,Shenj2007,Zhou2008,Gong2008,Shen2009,Tsoi2008,Huang2013,Yanc2014,%
Yan2014,Liao2015,Liao2016,Chenmt2017,Jiang2018,Zhou2020,Stolyarov2020,Zhao2020,Song2020,Fengs2021,Cai2021,Yin2022,Lih2023,Li2023}, single-photon router\cite{Zhoul2013,Zhang2013,Yanwb2014,Luj2014,Li2015,Yanepl2015,Cheng2016OE,Yangdc2018,Yan2018,Zhu2019,Ahumada2019,Huang2020,Cheng2021OE,Reny2022},
single-photon nonreciprocal transmission \cite{Reny2022,Shen2011,Xiaky2014,Sayrin2015,Xu2017,Yanwb2018,Wang2019,Yanch2020,Gu2022,Tang2022,Liu2022,Zhouj2023}, single-photon
transistor\cite{Chang2007,Witthaut2010,Neumeier2013,Gonzalez2016}, and so on.

In waveguide QED, the photons can be confined in the one-dimensional (1D)
waveguide and hence propagate in 1D space. It is feasible to couple the 1D
waveguide to the emitter. The emitter could be the atom (including the
artificial atom), cavity, atom-cavity hybrid structure, or other quantum
objects. In most practical operations, one can accurately manipulate the
location and number of the emitter. In the waveguide-emitter coupling
scheme, the transport of the guided single photon is affected by the
interaction between the guided photon and the emitter. For example, the
simplest and typical scheme is a two-level emitter coupled to a 1D
waveguide, in which the guided single photon moving towards the emitter with
a frequency on resonance is completely reflected due to the interference
between the reemitted wave and the incident wave \cite{2005Coherent,Shenj2007,Zhou2008}. When the 1D waveguide is coupled
to the emitter driven by the extra light field, the modulation of the extra
light field may constitute the control of the guided photon \cite{Gong2008,Shen2009,Witthaut2010,Zhoul2013,%
Yanc2014,Yan2014,Gu2022,Yanwb2018,Yan2018,Ahumada2019,Liu2022,Chang2007,Lih2023,Neumeier2013}. By coupling the
1D waveguide to one emitter or multiple emitters at multiple
coupling points, the transport of the guided
photon could also be influenced by the relative distance between the coupling points \cite{Tsoi2008,Liao2015,Liao2016,%
Chenmt2017,Zhao2020,Yin2022,Wang2019,Cheng2021OE,Fengs2021,Cai2021,Liu2022}.
Most studies focus on the single-photon transport affected by the
parameter such as the frequency, the light-matter interaction constant, or
other parameters relating to the light-matter interaction. The applications
of the quantum device would be affected by the role played by
thermodynamics. It will be valuable to realize the single-photon transport
controlled by the thermal parameter.

In this paper, we propose to realize that the singe-photon transport in a 1D
waveguide is controlled by the temperature. The waveguide is connected to a
thermal bath through a pair of intermediate interacting two-level atoms. The
temperature of the thermal bath has the impact on the transport of the guided
single photon. The single-photon reflection (or
transmission) coefficient can be controlled by
adjusting the temperature. It opens up a novel scenario for the control of
the single photon. Moreover, our scheme provides an approach for
implementing the optical thermometer, in which the temperature of the thermal
bath can be estimated by measuring the single-photon transport. It is known
that the measurement of temperature plays an important role in nature
science and technology. Especially, it is a non-trivial task to accurately determine an object's
temperature in the low-temperature region. It aroused
interest \cite{Mohammad2019,Vieira2023ExploringQT,HUANGFU2021127172,Seilmeier2014,Haupt2014,Hofer2017} with the increased capabilities of operating on small-scale
devices. Significantly, under certain conditions, a
very small change of the low temperature has the significant impact on
the transport of the guided photon in our scheme. It provides an avenue to precisely
measure the low temperature of the sample.

This paper is organized as follows. In Sec. \ref{modelandHamiltonian}, we
introduce the model and system Hamiltonian. In Sec. \ref%
{singlephotontransport}, we obtain the single-photon transport property with
the input-output formalism. In Sec. \ref{thermalcontrolandthermometer}, we
investigate the single-photon transport controlled by the temperature and
realize the thermometer sensitive in the low temperature region. In Sec. \ref%
{discussion}, the conclusion and discussion are made.

\section{Model and Hamiltonian}

\label{modelandHamiltonian}
\begin{figure}[t]
\includegraphics*[width=8cm, height=8cm]{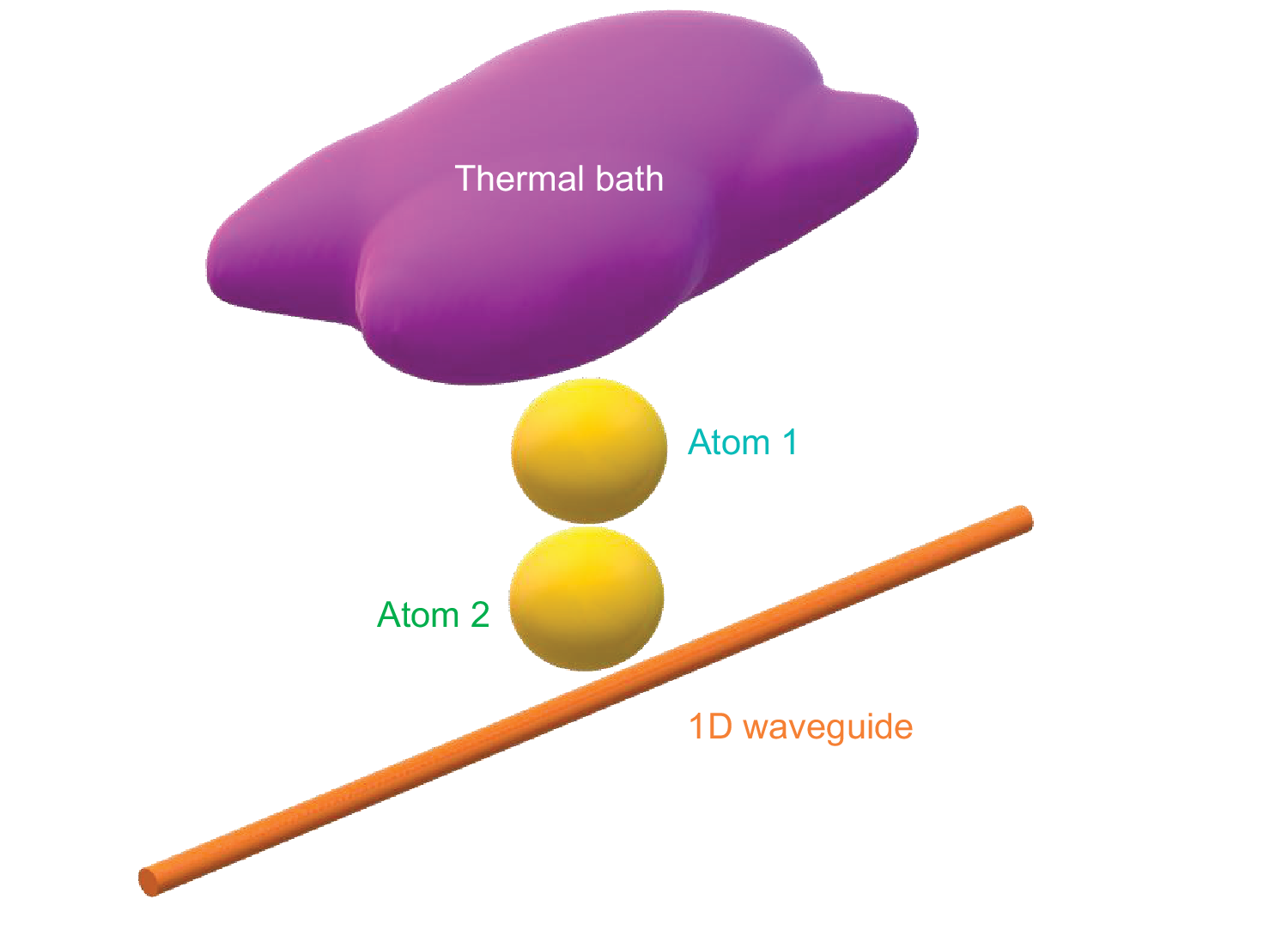}
\caption{Schematic diagram of the system under consideration. The atom 1
interacts with the atom 2. The atom 1 is linked to the thermal bath and the
atom 2 is coupled to a 1D waveguide.}
\label{Model}
\end{figure}
The system under consideration consists of two interacting two-level atoms,
a 1D waveguide, and a thermal bath, as shown in Fig. \ref{Model}. The two
atoms are labeled by atom $1$ and atom $2$, respectively. The atomic level
transition frequencies are denoted by $\omega _{m}$ ($m=1,2$). The atom $1$
is coupled to the thermal bath. The atom $2$ is coupled to the 1D waveguide.
The atomic Hamiltonian is%
\begin{equation}
H_{A}=\sum_{m=1,2}\frac{\omega _{m}}{2}\sigma _{m}^{z}+\frac{J}{2}\sigma
_{1}^{z}\sigma _{2}^{z}\text{,}  \label{HA}
\end{equation}%
where $J/2$ denotes the mutual interaction strength between the two atoms.
The operator $\sigma _{m}^{z}$ is the Pauli operator $\sigma _{z}$ that acts
on the atom $m$. The atomic mutual interaction does not exchange the
excitations between the two atoms. The free Hamiltonian of the guided photon
is \cite{2010Input}%
\begin{equation}
H_{W}=\int_{0}^{\infty }\omega r_{\omega }^{\dagger }r_{\omega }d\omega
+\int_{0}^{\infty }\omega l_{\omega }^{\dagger }l_{\omega }d\omega \text{,}
\label{HW}
\end{equation}%
The operator $r_{\omega }^{\dagger }$ ($l_{\omega }^{\dagger }$) creates a
right-propagating (left-propagating) guided photon with frequency $%
\omega $, and $r_{\omega }$ ($l_{\omega }$) is the corresponding photonic
annihilation operator. When the frequency of the guided photon is far away
from the cutoff frequency of the waveguide, the photonic dispersion relation
of the guided mode is approximately linear, i.e., $\omega
=v_{g}\left\vert k\right\vert $ with $k$ the wave number and $v_{g}$ the
photonic group velocity. Within the rotating-wave approximation, the
Hamiltonian denoting the waveguide-atom interaction is%
\begin{equation}
H_{A_{2}-W}=V\int_{0}^{\infty }\sigma _{2}^{ge}(r_{\omega }^{\dagger
}+l_{\omega }^{\dagger })d\omega +h.c.  \label{HAW}
\end{equation}%
where $V$ denotes the interaction strength between the atomic transition
dipole moment and the guided photonic field, and $\sigma
_{m}^{kl}=\left\vert k_{m}\right\rangle \left\langle l_{m}\right\vert $ ($%
k,l\in \{g,e\}$) is the atomic level transition and population operators.
The state vector $\left\vert g_{m}\right\rangle $ and $\left\vert
e_{m}\right\rangle $ represent the atom $m$'s ground and excited states,
respectively. The frequency width of the considered photonic pulse is much
smaller than its carrier frequency and the waveguide-atom coupling strength
varies little around the frequency. Therefore, the strength $V$ is
approximately frequency-independent. The thermal bath is considered as the
collection of decoupled quantum harmonic oscillators. The Hamiltonian of the
thermal bath is%
\begin{equation}
H_{B}=\sum_{\mathbf{p}}\omega _{\mathbf{p}}c_{\mathbf{p}}^{\dag }c_{\mathbf{p%
}},  \label{HB}
\end{equation}%
with $c_{\mathbf{p}}^{\dag }$ ($c_{\mathbf{p}}$) the bosonic creation
(annihilation) operator for the oscillation mode $\mathbf{p}$ with the frequency
$\omega _{\mathbf{p}}$. The interaction between the atom 1 and the thermal
bath is governed by the Hamiltonian%
\begin{equation}
H_{A_{1}-B}=\sigma _{1}^{x}\sum_{\mathbf{p}}\lambda _{\mathbf{p}}(c_{\mathbf{%
p}}^{\dag }+c_{\mathbf{p}})\text{.}  \label{HAB}
\end{equation}%
The operator $\sigma _{1}^{x}$ is the Pauli operator $\sigma _{x}$ that acts
on the atom $1$, and $\lambda _{\mathbf{p}}$ denotes the coupling strength
between the atom 1 and the oscillator mode $\mathbf{p}$ of the thermal bath.

\section{Single-photon transport}

\label{singlephotontransport} We consider that, initially, the waveguide is
in the vacuum state and the atom 2 is prepared in the ground state. The
atom-bath interaction has the impact on the excitation of the atom 1. The
thermal bath and the atom 1 are ultimately
thermal equilibrium. In this process, there is
no excitation transferred to the atom 2 and the waveguide stays its vacuum
state. The waveguide has no impact on the dynamics of the
subsystem composed by the thermal bath and the atoms. The system density operator can be
represented by the tensor product of density operator for the waveguide and
the density operator for the atom-bath subsystem. Therefore, during this
dynamics evolution process, we focus on the atom-bath subsystem, which is
governed by the Hamiltonian%
\begin{equation*}
H_{A-B}=H_{A}+H_{B}+H_{A_{1}-B}\text{.}
\end{equation*}%
In the weak-coupling regime, i.e., $J\ll \{\omega
_{1},\omega _{2}\}$, within the Born-Markov and secular approximations, one
can obtain the master equation as \cite{opensystem}%
\begin{eqnarray}
\dot{\rho} &=&-i[H_{A},\rho ]  \notag \\
&&+\frac{\mathcal{J(\omega }_{1})(n_{\mathcal{\omega }_{1}}+1)}{2}[2\sigma
_{1}^{ge}\rho \sigma _{1}^{eg}-\sigma _{1}^{ee}\rho -\rho \sigma _{1}^{ee}]
\notag \\
&&+\frac{\mathcal{J(\omega }_{1})n_{\mathcal{\omega }_{1}}}{2}[2\sigma
_{1}^{eg}\rho \sigma _{1}^{ge}-\sigma _{1}^{gg}\rho -\rho \sigma _{1}^{gg}]%
\text{.}  \label{ME}
\end{eqnarray}%
The density operator $\rho $ is the reduced density operator for the two
atoms, and $\mathcal{J(\cdot )}$ is the spectral function of the thermal
bath. The thermal bath is assumed ohmic, and the spectral function is linear
as $\mathcal{J(\omega }_{1})=\kappa \mathcal{\omega }_{1}$. The population
of the thermal bath's oscillator mode with frequency $\omega _{1}$ is given
by $n_{\omega _{1}}=(e^{\hbar \omega _{1}/k_{B}T}-1)^{-1}$,
which stems from the Bose-Einstein distribution. The symbol $k_{B}$ is the Boltzmann
constant and $T$ denotes the temperature of the thermal bath.

In the thermal equilibrium state, the reduced density operator for the atoms
is governed by the Eqn. (\ref{ME}) in the steady-state regime, defined as $%
\dot{\rho}_{ss}=0$. Consider the fact that\ $\left\langle
g_{1}e_{2}\right\vert \rho \left\vert g_{1}e_{2}\right\rangle =\left\langle
e_{1}e_{2}\right\vert \rho \left\vert e_{1}e_{2}\right\rangle =0$ and $%
Tr(\rho )=1$, the steady-state solution gives%
\begin{eqnarray}
\left\langle g_{1}g_{2}\right\vert \rho _{ss}\left\vert
g_{1}g_{2}\right\rangle &=&\frac{n_{\mathcal{\omega }_{1}}+1}{2n_{\mathcal{%
\omega }_{1}}+1},  \notag \\
\left\langle e_{1}g_{2}\right\vert \rho _{ss}\left\vert
e_{1}g_{2}\right\rangle &=&\frac{n_{\mathcal{\omega }_{1}}}{2n_{\mathcal{%
\omega }_{1}}+1}.  \label{SDEADYPHO}
\end{eqnarray}%
The effective temperature of the atom 1 is defined as $T_{1}^{eff}=\frac{%
\hbar\omega _{1}}{k_{B}}\left( \ln \frac{p_{g_1}}{p_{e_1}}\right) ^{-1}$,
where $p_{e_1}$ ($p_{g_1}$) denotes the probability of the atom 1 in
the state $\left\vert e\right\rangle $ ($\left\vert g\right\rangle $%
). It is easy to verify that the effective temperature of the atom 1 tends
to the thermal bath's temperature.

Then a photon is injected into the waveguide from the left port. The photon
would be absorbed and then reemitted by the atom 2. For convenience,
we bring in the even- and odd-parity modes as $a_{\mathcal{\omega }}=\frac{1%
}{\sqrt{2}}(r_{\mathcal{\omega }}+l_{\mathcal{\omega }})$ and $b_{\mathcal{%
\omega }}=\frac{1}{\sqrt{2}}(r_{\mathcal{\omega }}-l_{\mathcal{\omega }})$,
respectively. From the expressions of $H_{W}\ $and $H_{A_{2}-W}$, the atom
is coupled to the even mode, while the odd mode evolves freely in the
waveguide. Therefore, we focus on the evolution of the photon in the even
mode \cite{PhysRevA.76.062709}. The Heisenberg equations of the motion for $a_{\mathcal{\omega }}$ and $%
\sigma _{2}^{ge}$ are%
\begin{eqnarray}
\dot{a}_{\mathcal{\omega }} &=&-i\omega a_{\mathcal{\omega }}-ig\sigma
_{2}^{ge},  \notag \\
\dot{\sigma}_{2}^{ge} &=&-i\omega _{2}\sigma _{2}^{ge}-iJ\sigma
_{1}^{z}\sigma _{2}^{ge}+ig\int_{0}^{\infty }\sigma _{2}^{z}a_{\mathcal{%
\omega }}d\mathcal{\omega },  \label{HESE}
\end{eqnarray}%
with $g=\sqrt{2}V$. It follows that $\sigma _{2}^{ge}$ has the direct impact
on the time evolution of $a_{\mathcal{\omega }}$, and $\sigma _{1}^{z}$ has
the direct impact on the evolution of $\sigma _{2}^{ge}$. Phenomenally, the
waveguide-atom interaction does not affect the excitations of the atom 1 and
the thermal bath because the atomic mutual interaction does not exchange the
atomic excitations. This can be verified by $[\sigma _{1}^{z},H_{A-W}]=[c_{%
\mathbf{p}}^{\dag }c_{\mathbf{p}},H_{A-W}]=0$, with%
\begin{equation*}
H_{A-W}=H_{A}+H_{W}+H_{A_{2}-W}.
\end{equation*}%
We consider that the atom 1 and the thermal bath stay thermal
equilibrium and take $\sigma _{1}^{z}=\frac{1-e^{\hbar \omega
_{1}/k_{B}T}}{1+e^{\hbar \omega _{1}/k_{B}T}}$.

\begin{figure}[t]
\includegraphics*[width=8cm, height=5cm]{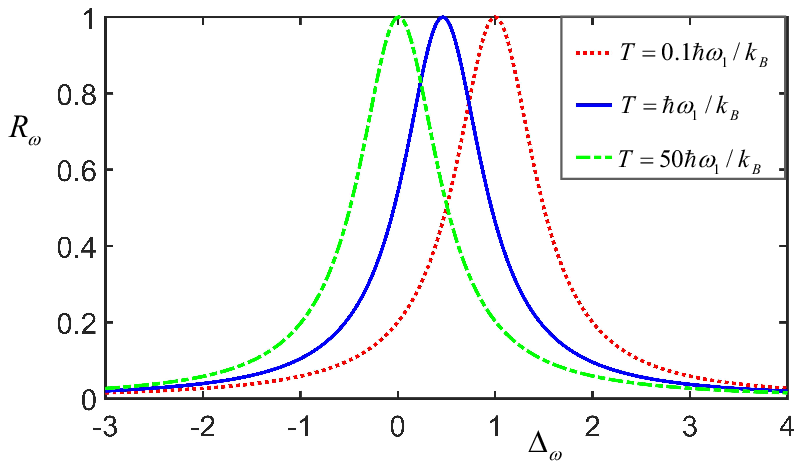}
\caption{The single-photon reflection probability against the detuning $%
\Delta _{\protect\omega }$. The red dotted, blue solid, and green dashed
dotted lines denote the cases when $T=\frac{0.1\hbar \protect\omega _{1}}{%
k_{B}}$, $T=\frac{\hbar \protect\omega _{1}}{k_{B}}$, and $T=\frac{50\hbar
\protect\omega _{1}}{k_{B}}$, respectively. The parameters $\Delta _{\protect%
\omega }$ and $\protect\gamma $ are in units of $J$ and $\protect\gamma =0.5$%
.}
\label{Rvsdelta}
\end{figure}
The single-photon transport property can be obtained with the input-output formalism \cite{walls}.
By performing the standard procedure, the input-output relation is found as%
\begin{equation}
a_{out}(t)=a_{in}(t)-i\sqrt{2\gamma }\sigma _{2}^{ge}(t),  \label{IN-OUT}
\end{equation}%
where $\gamma =\pi g^{2}$ and the input and output operators are defined as
\begin{eqnarray*}
a_{in}(t) &=&\frac{1}{\sqrt{2\pi }}\int_{-\infty }^{\infty }d\mathcal{\omega
}a_{\mathcal{\omega }}(t_{0})e^{-i\mathcal{\omega }(t-t_{0})}, \\
a_{out}(t) &=&\frac{1}{\sqrt{2\pi }}\int_{-\infty }^{\infty }d\mathcal{%
\omega }a_{\mathcal{\omega }}(t_{1})e^{-i\mathcal{\omega }(t-t_{1})}.
\end{eqnarray*}%
In the integration, the lower bound of the frequency $\omega $ is
approximately extended from $0$ to $-\infty $. The Fourier transformations
of $a_{in}^{\dagger }(t)$, $a_{out}^{\dagger }(t)$, and $\sigma _{2}^{eg}(t)$
are denoted by $a_{in}^{\dagger }(\omega )$, $a_{out}^{\dagger }(\omega )$,
and $\sigma _{2}^{eg}(\omega )$, respectively. Here we consider the
so-called \textit{weak excitation limit}, in which the atom 2 is mostly in
the ground state and hence $\sigma _{2}^{z}$ is approximately $-1$. The
similar approximations have been utilized in many quantum optics
calculations. For the guided photons scattered by the emitter, the
connection between the scattering theory and the input-output formalism has
been found \cite{2010Input}. In Ref. \cite{2010Input}, the single-photon scattering property
is obtained with the input-output formalism in
the general case, in which the \textit{weak excitation limit} is not considered.
It shows that the single-photon transport in a 1D waveguide coupled to a two-level emitter obtained
in the \textit{weak excitation limit} coincides with the outcomes obtained beyond the \textit{weak excitation limit}.
For a two-level atom inverted by the guided unidirectional single-photon pulse,
the atomic inversion can maximally reach half at particular time points only for a unique pulse \cite{PhysRevA.82.033804}.
The \textit{weak excitation limit} is enough for our scheme. Then, one
can obtain%
\begin{equation}
a_{out}(\omega )=\bar{t}_{\omega }a_{in}(\omega ),  \notag
\end{equation}%
with%
\begin{equation}
\bar{t}_{\omega }=\frac{\Delta _{\omega}-J+i\gamma +(\Delta _{\omega }+J+i\gamma
)e^{-\hbar \omega _{1}/k_{B}T}}{\Delta _{\omega}-J-i\gamma +(\Delta _{\omega
}+J-i\gamma )e^{-\hbar \omega _{1}/k_{B}T}}  \label{TRANS_E}
\end{equation}%
and $\Delta _{\omega }=\omega _{2}-\omega $ being the detuning between the atom 2
and the guided photon with the frequency $\omega$. Due to the
atom-waveguide interaction, the output photonic even mode gains a
temperature-dependent phase shift.

Going back to the right-moving and left-moving modes, the transmission and
reflection amplitudes of the guided single photon with frequency $\omega $
are%
\begin{eqnarray}
t_{\omega } &=&\frac{(\Delta _{\omega }-J)+(\Delta _{\omega }+J)e^{-\hbar
\omega _{1}/k_{B}T}}{(\Delta _{\omega }-J-i\gamma )+(\Delta _{\omega
}+J-i\gamma )e^{-\hbar \omega _{1}/k_{B}T}},  \notag \\
r_{\omega } &=&\frac{i\gamma +i\gamma e^{-\hbar \omega _{1}/k_{B}T}}{(\Delta
_{\omega }-J-i\gamma )+(\Delta _{\omega }+J-i\gamma )e^{-\hbar \omega
_{1}/k_{B}T}}.  \label{TR}
\end{eqnarray}%
When the atom 2 is decoupled to the atom 1, i.e., $J=0$, the scheme is simplified into a
1D waveguide coupled to the atom 2 and the amplitudes become $t_{\omega }=\frac{%
\Delta _{\omega }}{\Delta _{\omega }-i\gamma }$ and $r_{\mathcal{\omega }}=%
\frac{i\gamma }{\Delta _{\mathcal{\omega }}-i\gamma }$, which agree with the
results obtained in a 1D waveguide coupled to a two-level emitter \cite{2005Coherent}. When $%
T\rightarrow 0$, the amplitudes turn out to be $t_{\mathcal{\omega }%
}\rightarrow \frac{\omega _{2}-J-\omega}{\omega _{2}-J-\omega-i\gamma }$ and $%
r_{\omega }\rightarrow \frac{i\gamma }{\omega _{2}-J-\omega-i\gamma }$, which
agree with the case that the guided photon is scattered by a two-level atom
with the transition frequency $\omega _{2}-J$. This can be understood by the
fact that the four eigenstates of the atomic Hamiltonian (\ref{HA}) are $%
\left\vert g_{1}g_{2}\right\rangle $, $\left\vert g_{1}e_{2}\right\rangle $,
$\left\vert e_{1}g_{2}\right\rangle $, and $\left\vert
e_{1}e_{2}\right\rangle $, and the four corresponding eigenvalues are $\frac{%
-\omega _{1}-\omega _{2}+J}{2}$, $\frac{-\omega _{1}+\omega _{2}-J}{2}$, $%
\frac{\omega _{1}-\omega _{2}-J}{2}$, and $\frac{\omega _{1}+\omega _{2}+J}{2%
}$, respectively. The interaction between the waveguide and the atom 2
drives the atomic transitions $\left\vert g_{1}g_{2}\right\rangle
\leftrightarrow \left\vert g_{1}e_{2}\right\rangle $ and $\left\vert
e_{1}g_{2}\right\rangle \leftrightarrow \left\vert e_{1}e_{2}\right\rangle $%
. At $T\rightarrow 0$, the atoms are in the pure state $\left\vert
g_{1}g_{2}\right\rangle $ at the moment when the photon is injected into the
waveguide. In this case, only the transition $\left\vert g_{1}g_{2}\right\rangle
\leftrightarrow \left\vert g_{1}e_{2}\right\rangle $ participates in the
dynamic processes and hence it can be considered that the waveguide is
coupled to a two-level atom.
As $T$ increases, the pure state $\left\vert
g_{1}g_{2}\right\rangle $ at thermal equilibrium turns out to be a mixed state
and hence the transition $\left\vert
e_{1}g_{2}\right\rangle \leftrightarrow \left\vert e_{1}e_{2}\right\rangle $
has the impact on single-photon transport.
It follows that the terms multiplied
by $e^{-\hbar \omega _{1}/k_{B}T}$ in Eqns. (\ref{TR}) stem from the transition $\left\vert
e_{1}g_{2}\right\rangle \leftrightarrow \left\vert e_{1}e_{2}\right\rangle $
and the other terms stem from the transition $\left\vert
g_{1}g_{2}\right\rangle \leftrightarrow \left\vert g_{1}e_{2}\right\rangle $.
This can also be understood by the fact that the detuning between the guided photon and the transition $%
\left\vert g_{1}g_{2}\right\rangle \leftrightarrow \left\vert
g_{1}e_{2}\right\rangle $ ($\left\vert e_{1}g_{2}\right\rangle
\leftrightarrow \left\vert e_{1}e_{2}\right\rangle $) is $%
\Delta _{\omega}-J$ ($\Delta _{\omega}+J$).

The transmission and reflection probabilities of the single photon with frequency $\omega$ are defined as $%
T_{\omega }=\left\vert t_{\omega }\right\vert ^{2}$ and $R_{\omega
}=\left\vert r_{\omega }\right\vert ^{2}$. In Fig. \ref{Rvsdelta}, we
simulate the single-photon reflection probability against $\Delta _{\omega}$
for different temperatures. The point at which $R_{\omega }$ reaches its
maximum value shifts as the temperature $T$ varies. Therefore, the
temperature has the significant impact on the single-photon reflection
spectrum.

\section{Thermal control of single-photon transport and optical thermometer}

\label{thermalcontrolandthermometer}
\begin{figure}[t]
\includegraphics*[width=8cm, height=5.3cm]{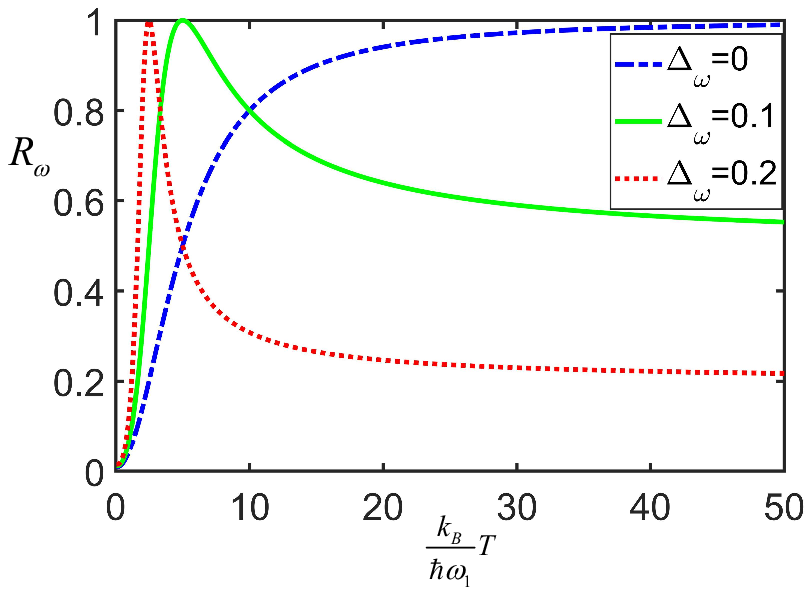}
\caption{The single-photon reflection probability against the thermal bath's
temperature. The blue dashed dotted, green solid, and red dotted lines
denote the cases when $\Delta _{\protect\omega }=0$, $\Delta _{\protect%
\omega }=0.1$, and $\Delta _{\protect\omega }=0.2$, respectively. The
parameters $\Delta _{\protect\omega }$ and $\protect\gamma $ are in units of
J and $\protect\gamma =0.1$.}
\label{RvsT}
\end{figure}
It is feasible to realize two implementations. One implementation is to control
the single-photon transport by the temperature of the thermal bath. The
other implementation is to achieve the thermometer sensitive to the slight
changes of the low temperatures. In the thermometer, the temperature of the
thermal bath is estimated by measuring the single-photon transport.

By introducing $\Gamma =\frac{\gamma }{J}$, $x=\frac{1-e^{\hbar \omega
_{1}/k_{B}T}}{1+e^{\hbar \omega _{1}/k_{B}T}}$, and $x_{0}=-\frac{\Delta
_{\omega }}{J}$, the single-photon transmission and reflection probabilities
are written as%
\begin{eqnarray}
T_{\omega } &=&\frac{(x-x_{0})^{2}}{(x-x_{0})^{2}+\Gamma ^{2}}  \notag \\
R_{\omega } &=&\frac{\Gamma ^{2}}{(x-x_{0})^{2}+\Gamma ^{2}},  \label{PRO}
\end{eqnarray}%
It is obvious that $T_{\omega }+R_{\omega }=1$. There are similarities
between the lineshape of $R_{\omega }$ against $x$ and the lineshape of the
Lorentz spectrum. At $x=x_{0}$, $R_{\omega }$ reaches its maximum value $1$,
the single photon is completely reflected. In the region $x<x_{0}$ ($x>x_{0}$%
), $P_{r}$ monotonically increases (decreases) as $x$ increases. In the
region $(x-x_{0})^{2}\gg \Gamma ^{2}$, the single-photon transmission
probability is near unity. At $\left\vert x-x_{0}\right\vert =\Gamma $, $%
R_{\omega }$ reaches half of its maximum value. When the value of $\Gamma $
is small, $R_{\omega }$ rapidly varies with $x$ in the region of $x$ near $%
x_{0}\pm \Gamma $. The value of $x$ increases with the increasing $T$. The
lower and upper bounds of $x$ are $-1$ and $0$, respectively. Therefore,
when $\Delta _{\omega }=0$, $R_{\omega }$ monotonically increases with the
increasing $T$ and tends to unity when $T\ $is large enough. When $\Delta
_{\omega }=J$, $R_{\omega }$ reaches the maximum value of 1 at $T=0$ and
monotonically decreases with the increasing $T$. When $\Delta _{\omega
}\in \lbrack 0,J]$, $R_{\omega }$ can reach $1$ by appropriately setting the
temperature $T$. However, when $\Delta _{\omega }$ is outside the range $[0$%
, $J]$, $R_{\omega }$ is smaller than $1$ for any temperature. To well control the single-photon transport, the detuning $\Delta
_{\omega }$ should be inside the region $[0,J]$ so that the single photon
can be completely reflected at an appropriate temperature. Moreover, the value
of $\gamma $ is limited\ so that the reflection probability can be near null
at the appropriate temperature. In Fig. \ref{RvsT}, we plot the single-photon
reflection probability $R_{\omega }$ against the temperature $T$ for
different detunings. The single-photon reflection probability is well
controlled by adjusting the temperature. Therefore, the thermal control of
the single-photon transport is realized.

\begin{figure}[t]
\includegraphics*[width=8cm, height=11cm]{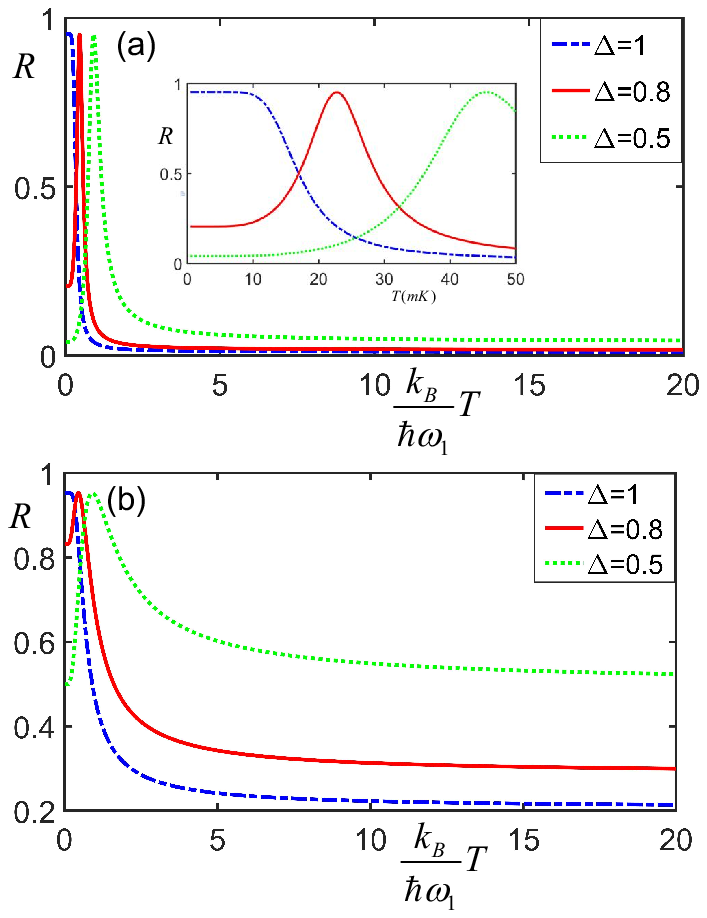}
\caption{Numerical simulation of the single-photon-pulse reflection
probability against the temperature. The blue dashed dotted, red solid, and
green dotted lines denote the cases when $\Delta _{\protect\omega }=1$, $%
\Delta _{\protect\omega }=0.8$, and $\Delta _{\protect\omega }=0.5$,
respectively. In (a) and (b), we take $\protect\gamma =0.1$ and $\protect%
\gamma =0.5$, respectively. The parameters $\protect%
\gamma $ and $\Delta _{\protect\omega }$ are in units of J
and $\protect\eta =\gamma/20$. The insert
panel in (a) shows the details of the corresponding three lines when $\omega_{1}$ is assigned the value of $2\pi\times 1GHz$.}
\label{PlusRvsT}
\end{figure}
We proceed to realize the thermometer. The waveguide and the two atoms
constitute the thermometer. The atom 1 plays the role of the thermometer's
probe. We consider that the temperature $T$ is unknown and the
thermometer's probe is not coupled to the thermal bath before measuring
the temperature. The approach to estimate the temperature can be divided
into two steps. In step one, the thermometer's probe is linked to the
thermal bath. In step two, the single-photon pulse is injected into
the waveguide after the steady state is realized. Then by detecting the photonic
transmission or reflection probability, the value of $T$ is estimated based
on the relation between the temperature and the single-photon transport
property. In Fig. \ref{PlusRvsT}, we numerically simulate the reflection
probability of the single-photon pulse with a Lorentzian frequency
distribution against the temperature. The Lorentzian spectrum has the form
of $\frac{\sqrt{\eta /\pi }}{i(\omega _{t}-\omega )+\eta }$, with $\omega
_{t}$ the carrier frequency of the pulse and $\eta $ the spectral width. The
detuning $\Delta$ in Fig. \ref{PlusRvsT} is the difference
between $\omega _{2}$ and the carrier frequency. The reflection probability
is labeled by $R$ instead of $R_{\omega }$ because it has integrated $\omega
$. In Fig. \ref{PlusRvsT}, $R$ is sensitive to the small change of $\frac{k_{B}}{\hbar \omega
_{1}}T$ when the value of $\frac{k_{B}}{\hbar \omega _{1}}T$ is small. Therefore, for
the appropriate value of $\omega _{1}$, it is feasible
that the single-photon transport is significantly sensitive to a very small change of $T$ in the low-temperature region. When the resonant frequency of the atom
1 is in the microwave domain as $2\pi \times 8.136GHz$ \cite{Coherentcoupling}, one can obtain $%
\frac{k_{B}}{\hbar \omega _{1}}\sim 2.56K^{-1}$. When the resonant frequency
of the atom 1 is $2\pi \times 1GHz$ \cite{GU20171}, the value of $\frac{k_{B}}{\hbar \omega
_{1}}$ is near $\frac{1}{50}mK^{-1}$. In this case, as shown by the insert
panel in Fig. \ref{PlusRvsT} (a), the single-photon transport reflection obtains a
obvious change when the temperature varies in the order of $10^{-3}K$. The blue dashed dotted
lines represent the case when $\Delta=J$, in which the reflection
probability monotonically changes with $T$ as discussed above. For any
value of $R$, there is only one corresponding temperature. By detecting the
transport of the pulse, the temperature can be estimated. However, the red
solid and green dotted lines do not monotonically change with $T$. For a
given value of $R$, there may be two corresponding temperatures. In this case,
the temperature can be estimated by respectively detecting the transports of
the pulses with different carrier frequencies. In the thermometer, it is not
necessary that the minimum and maximum values of $R$ are near null and unity,
respectively. The key is that the single-photon transport changes significantly
against the temperature to distinguish the small change of the temperature.

\section{Discussions}

\label{discussion} We investigate the transport of the single photon in a 1D
waveguide coupled to a two-level atom, which interacts with another atom
linked to a thermal bath. The single-photon transmission and reflection
coefficients are obtained with input-output formalism. The outcomes show
that the single-photon transport can be controlled by the temperature of the
thermal bath. Moreover, the atom-waveguide system constitutes a thermometer
sensitive in the low temperature region. In practice, the 1D waveguide
coupled to the atoms can be realized in several nanoscale schemes, such as
quantum dots coupled to the surface plasmons or line defects in photonic
crystals, quantum dots or nanocrystals coupled to semiconductor or diamond nanowire,
superconducting qubits coupled to transmission lines, and so on. For the two
atoms realized by two superconducting qubits, the desired atomic mutual
interaction, which does not transfer the excitation from one qubit to
the other, can be realized with a superconducting quantum interference device \cite{Neumeier2013}.
The thermal bath could be composed by the transmission line with the guided
photons in the thermal state.

\bibliographystyle{apsrev4-2}

\end{document}